\documentclass[12pt,onecolumn,draftclsnofoot]{IEEEtran}

\usepackage{cite}
\usepackage{amsmath,amssymb,amsfonts}
\usepackage{algorithmic}
\usepackage{graphicx}
\usepackage{textcomp}
\usepackage{booktabs}
\usepackage{makecell}
\usepackage{subcaption}

\usepackage{algorithmic}
\usepackage{graphicx}
\usepackage{textcomp}
\usepackage{booktabs}
\usepackage{makecell}
\usepackage{subcaption}
\def\BibTeX{{\rm B\kern-.05em{\sc i\kern-.025em b}\kern-.08em
    T\kern-.1667em\lower.7ex\hbox{E}\kern-.125emX}}

\ifCLASSINFOpdf
\else
\fi
\begin{document}
%
\title{Computationally Efficient Machine-Learning-Based Online Battery State of Health Estimation}
%
%
%

\author{Abhijit Kulkarni, \IEEEmembership{Member, IEEE}, and Remus Teodorescu \IEEEmembership{Fellow, IEEE}
\thanks{This work was supported in part by the the OPENSRUM grant (223286)
from MSCA-IF European Commission and  Smart Battery Villum Investigator Grant (222860) from VILLUM FONDEN.}
\thanks{The authors are with the Department of Energy, Aalborg University, Denmark. }}

%
%

\markboth{}{}
%



\maketitle

\begin{abstract}
A key function of  battery management systems (BMS) in e-mobility applications is estimating the battery state of health (SoH) with high accuracy. This is typically achieved in commercial BMS using model-based methods. There has been considerable research in developing data-driven methods for improving the accuracy of SoH estimation. The data-driven methods are diverse and use different machine-learning (ML) or artificial intelligence (AI) based techniques. Complex AI/ML techniques are difficult to implement in low-cost microcontrollers used in BMS due to the extensive use of non-linear functions and large matrix operations. This paper proposes a computationally efficient and data-lightweight SoH estimation technique. Online impedance at four discrete frequencies is evaluated to derive the features of a linear regression problem. The proposed solution avoids complex mathematical operations and it is well-suited for online implementation in a commercial BMS.  The accuracy of this method is validated on two experimental datasets and is shown to have a mean absolute error (MAE) of less than 2\% across diverse training and testing data.
\end{abstract}


%
\IEEEpeerreviewmaketitle

\section{Introduction}
\label{sec:introduction}
\IEEEPARstart{B}{attery} management systems (BMS) are indispensable to ensure safe and optimal performance from the battery packs used for e-mobility applications \cite{bms_high_level}. The passenger electric vehicles are increasing significantly and will help to reduce the $CO_2$ emissions globally. In addition, there has been a significant growth in the all-electric aviation \cite{lib_aero_demand_elsv}. Electric vertical takeoff and landing (eVTOL) aircraft are actively developed worldwide in order to reduce the traffic congestion and improve mobility in urban environment. For aerospace applications, it is very important that the BMS has very high accuracy in battery state estimations \cite{evtol_soh} such as state of charge (SoC), state of health (SoH) and state of temperature (SoT). Aerospace applications require redundancy to ensure safety. Thus, in addition to the hardware redundancy, it is important to have different approaches to estimate the states and provide a more accurate result. 

Typical state estimations by commercial BMSs involve model-based methods. Cell voltage, current and temperatures are used to arrive at  cell and pack-level states using model-based calculations \cite{plett_vol1}. Since Li-ion cells are very complex to model completely analytically, recently data-driven approaches are becoming very popular. The state-of-the-art on the data driven SoH estimation methods has been covered very well in multiple review articles \cite{soh_gen_rev1,soh_gen_rev2,soh_gen_rev3}. The data-driven methods are quite diverse and use different types of inputs from the BMS for estimating the SoH. As indicated in \cite{yunhong_soh} the data-driven methods typically do not use any model. These methods use BMS sensed signals such as cell voltage, current and temperature along with an AI/ML method which can be based on deep neural networks (DNN) \cite{dnn_generic1}, support vector regression (SVR) \cite{soh_svr1}, long short-time memory neural networks (LSTM) \cite{lstm_gpr1}, etc. Typically, the data-driven methods have less explainability due to the black-box nature of the AI/ML technique used. These methods may also require complex calculations since they may involve neural networks with non-linear activation functions. The high computational complexity makes them suitable to be implemented in cloud \cite{yunhong_soh,cloud1,cloud2}. Thus, the accuracy of data-driven approach can be used along with cloud computing. 

Model-based methods  use analytical models to represent Li-ion cells, whose parameters can be used as an indication of the aging or SoH in the cell. There are many model based methods reported in literature that use Kalman filters \cite{kalman1,kalman2}.  Commonly used model for cells is the equivalent circuit model (ECM) that consists of resistors, capacitors and non-linear elements such as Warburg element and/or constant phase elements (CPEs) \cite{plett_vol1}. Some works use electrochemical models such as single-particle model or P2D model for estimating SoH \cite{em_spm1,em_weihan}. ECM based models are more popular due to their simplicity, better interpretation of the model parameters and low computational efforts. The most comprehensive method to obtain the impedance information and hence develop the ECM is electrochemical impedance spectroscopy (EIS). In EIS, small-signal voltage or currents are applied to a cell and impedance is measured at different frequencies \cite{eis_soh_mona}. True-EIS is difficult to implement online since a controller with a very high bandwidth and high accuracy is needed \cite{eis_limitation_dis}. Thus, there are alternate methods reported in literature that determine the battery impedance. For example, in \cite{eis_limitation_dis,prbs1,prbs2} band-limited time domain signals such as pseudorandom binary sequence (PRBS) are used. These methods have the advantage of getting the ECM parameters in a shorter time. However, the methods in literature typically use optimization methods \cite{eis_optimization1,eis_optimization2,ak_arxiv} to fit the parameters. It is difficult to implement the optimization routines in online BMS. Nevertheless, many works use the ECM parameters and data-drive methods for estimating the SoH.

In \cite{cnn_eis}, convolutional neural networks (CNN) are used to determine SoH from EIS data. Gaussian process regression (GPR) is used in \cite{eis_soh_mona} on a comprehensive dataset generated in-house. GPR is a computationally complex method due to large matrix operations and matrix inversions. The research article \cite{eis_good_summary} provides a good summary of different AI/ML techniques and feature extraction methods used based EIS data. All these methods are strongly connected to the availability of EIS data, which is difficult to obtain in-field.

In this paper, a low-complexity and datalight approach is utilized for determining the SoH. In this method, only four discrete impedance values are needed as inputs. Using these impedance values, an ECM is developed analytically without making use of complex optimization methods. 
Once the ECM parameters are known, linear regression from machine learning is used to estimate the SoH. The proposed approach differs from the available ECM-based or EIS-based methods because:
\begin{itemize}
    \item Proposed method does not need EIS data. Instead four discrete impedance values are needed.
    \item Proposed method does not use nonlinear or complex curve-fit methods to determine the ECM parameters from the measured impedances.
    \item Proposed method uses linear regression, which can be easily deployed in any low-cost BMS processor after training.
\end{itemize}
The discrete impedance values can be obtained either as small-signal impedances or as operando impedances \cite{operando_eis}. Since the Nyquist plots of operando and EIS impedances have similar trends, the proposed approach works on both. Thus, it is also well-suited for online impelementation where the BMS can measure four discrete impedances without the need of extensive hardware modifications.

The proposed methodology has been validated using two separate datasets \cite{german_dataset, wmg_dataset}. It is shown that the RMSE and MAE are under 2\% and the developed model has a very good $R^2$ score for various scenarios.

\section{Model Development with Online Parameter Extraction}
\begin{figure}[htbp]
    \centering
    \includegraphics[width=0.95\linewidth]{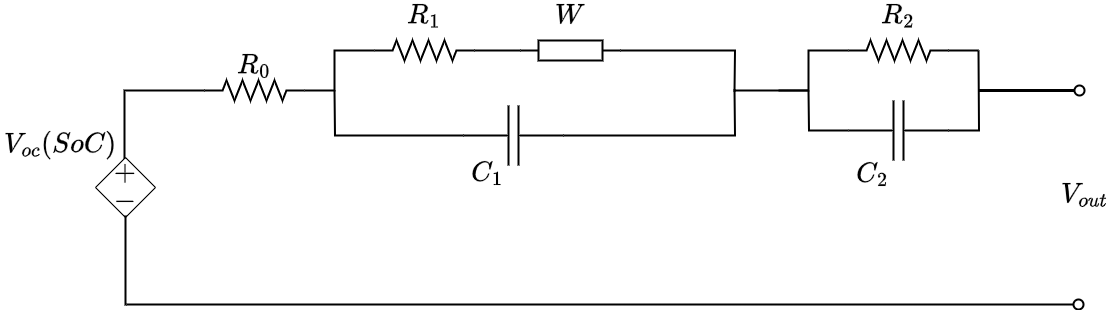}
    \caption{Equivalent circuit model (ECM) of a Li-ion cell considered in this paper.}
    \label{fig:ecm}
\end{figure}

In this paper, the ECM shown in Fig. \ref{fig:ecm} is used. This consists of the electrolytic resistance $R_0$, Randles's circuit with $R_1$, $C_1$, Warburg element $W$ and an additional RC branch with $R_2$ and $C_2$. The Warburg impedance is defined as,
\begin{equation}
    W = \frac{A_w}{\sqrt{j\omega}} \label{eq:warburg}
\end{equation}
In (\ref{eq:warburg}), Warburg impedance W is a function of frequency $\omega$ in rad/s. Its amplitude is affected by the gain term $A_w$, while it always gives a phase shift of $45^\circ$.

The parameters of the ECM are typically determined using offline methods, thus it is difficult to use them for online state estimation. In this work, a new method is proposed that enables the online estimation of the ECM parameters, which are then used to estimate the SoH using a ML based method. The definition of SoH used in this paper is based on the cell capacity as indicated in (\ref{eq:soh_def}).
\begin{equation}
    SoH = \frac{Q_{max}}{Q_{rated}} \label{eq:soh_def}
\end{equation}
In (\ref{eq:soh_def}), $Q_{rated}$ is the rated capacity (in Ah) for the cell under consideration and $Q_{max}$ is the maximum available capacity (in Ah) at the time of measurement. This definition is preferred compared to the definition based on the internal resistance \cite{soh_defn1} since for automotive applications, a measure of current capacity can be connected to the driving range directly. 

In this paper, the ECM parameters in Fig. \ref{fig:ecm} are related to the SoH using linear regression approach. The feature set used for the linear regression are represented as,
\begin{equation}
    \boldsymbol{x} = [R_0, ~R_1, ~R_2, ~A_w, ~C_1, ~C_2]^T \label{eq:regr_freature}
\end{equation}
Thus, the SoH estimation approach in this paper is expressed as follows in (\ref{eq:regr_formulation}),
\begin{equation}
    \hat{SoH} = \boldsymbol{\beta}\boldsymbol{x} + \beta_0 \label{eq:regr_formulation}
\end{equation}
The regression parameter vectors $\boldsymbol{\beta}$ and the offset $\beta_0$ are determined using the training process. If the feature vector is known at a given instant of time, the SoH can be estimated using (\ref{eq:regr_formulation}) online. Thus, it is important to develop an online method that can estimate the ECM parameters. This is described in the following subsection.

\subsection{Online estimation of ECM parameters}
In \cite{ak_arxiv}, analytical expressions are derived to determine the ECM parameters using three discrete impedance values since it uses a simplified model. The ECM used in \cite{ak_arxiv} differs from the ECM in Fig. \ref{fig:ecm} by the additional RC branch with $R_2$ and $C_2$. In this paper, this additional branch is used to have more features for the SoH estimation and for improving the overall accuracy for SoH estimation. Since, additional parameters are used, the method reported in \cite{ak_arxiv} needs to be enhanced. Thus, in this paper, four discrete impedances of the Li-ion cell are determined. Using these discrete impedances, analytical expressions are derived to determine the full ECM parameter set or the feature set in (\ref{eq:regr_freature}).

\begin{figure}
    \centering
    \includegraphics[width=0.5\linewidth]{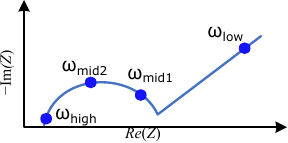}
    \caption{Four discrete frequencies labeled on an EIS Nyquist plot which are used to obtain the impedance vector.}
    \label{fig:freq_sel}
\end{figure}

\begin{figure}
    \centering
    \includegraphics[width=0.95\linewidth]{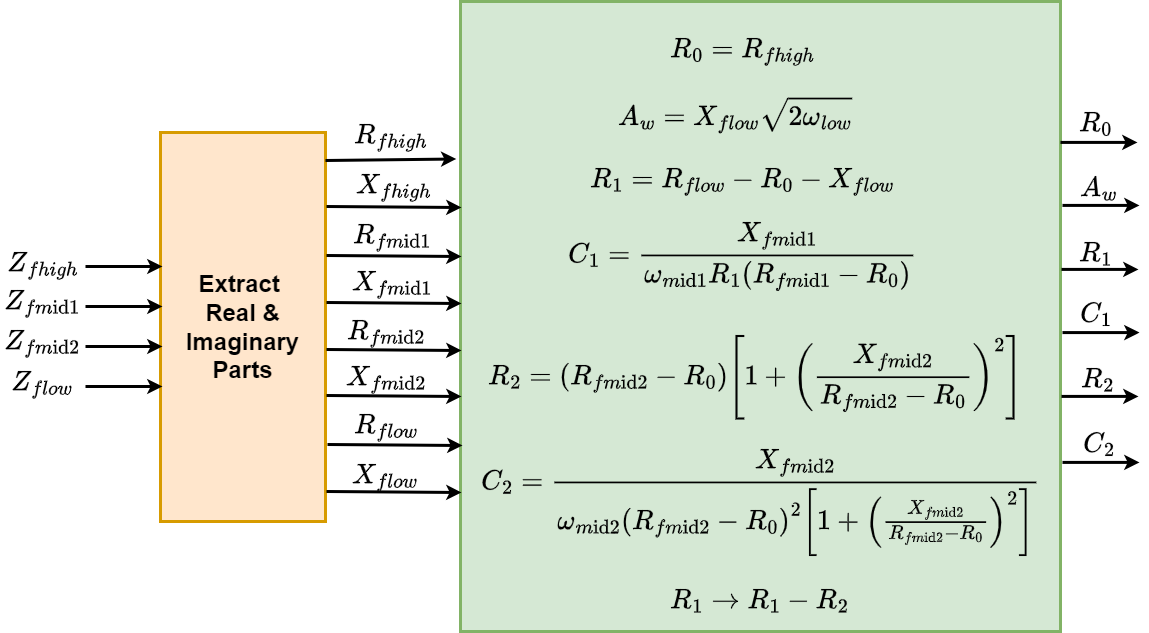}
    \caption{Analytical expressions developed to determine ECM parameters from four discrete impedance values.}
    \label{fig:ecm_param_extract}
\end{figure}

The four discrete impedance values used in the proposed approach are represented in the following impedance vector.
\begin{equation}
    \boldsymbol{Z} = [Z_{fhigh},~Z_{fmid1},~Z_{fmid2},~Z_{flow}]^T \label{eq:imp_vector}
\end{equation}
The selection of the impedances is mainly based on the frequencies at which impedance is evaluated. The frequencies are chosen to be well separated so that they are at least a decade away from each other. An illustration of the impedances needed in the proposed method is shown graphically in Fig. \ref{fig:freq_sel} on a Nyquist plot.
The impedance vector in (\ref{eq:imp_vector}) corresponds to the frequency vector in (\ref{eq:freq_vector}), which are also marked in Fig. \ref{fig:freq_sel}.
\begin{equation}
    \boldsymbol{\omega} = [\omega_{high},~\omega_{mid1},~\omega_{mid2},~\omega_{low}]^T \label{eq:freq_vector}
\end{equation}
The impedances in (\ref{eq:imp_vector}) are complex numbers consisting of real and imaginary parts. Thus, at four frequencies corresponding to the frequency vector (\ref{eq:freq_vector}), there are eight unique impedance elements that can be used to derive the ECM parameter values. The derivation is provided in Appendix I. The summary of the equations is shown graphically in Fig. \ref{fig:ecm_param_extract}.

\subsection{Online extraction of cell impedance}
\begin{figure*}[htbp]
    \centering
    \includegraphics[width=0.8\linewidth]{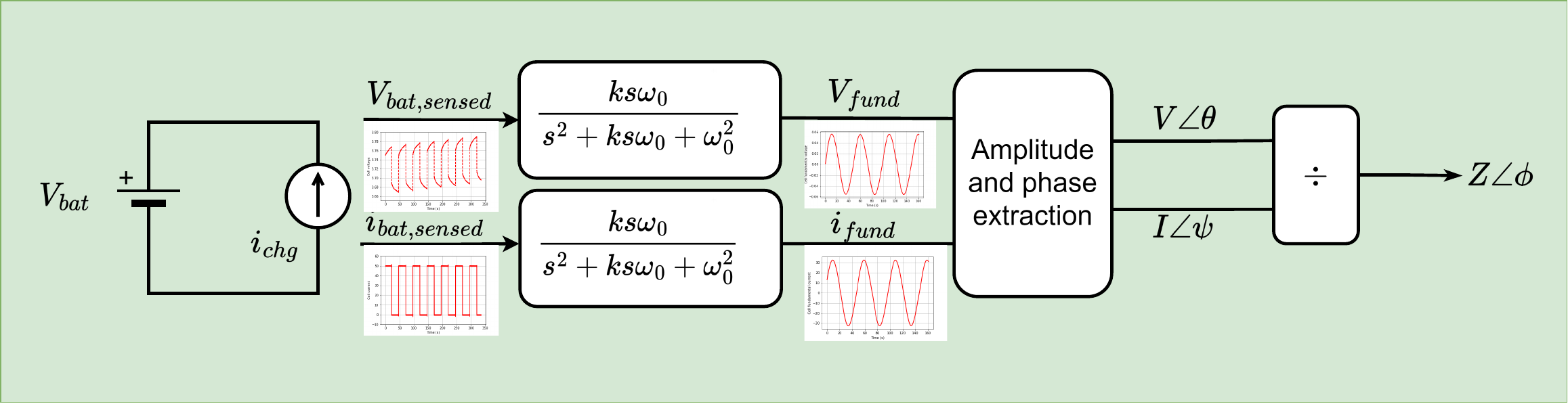}
    \caption{Low-complexity method to estimate cell impedances online.}
    \label{fig:online_imp_calc}
\end{figure*}

In order to extract the four impedances, the cell can be excited by pulsed currents at the desired frequencies. Cell voltage and current are then passed through bandpass filters to extract the fundamental voltage and currents. Then impedance can be determined using the extracted sinusoidal components. This approach is shown in Fig. \ref{fig:online_imp_calc}.
This approach does not require any complex operations such as fast Fourier transform (FFT). Simple second order filters are used to extract the fundamental components. Thus, the impedance extraction is computationally simple. Once the impedance vector is known, the ECM parameters can be extracted using the algebraic expressions shown in Fig. \ref{fig:ecm_param_extract}. It can be observed that the proposed method does not use any nonlinear optimization method and simple algebraic expressions are used to achieve high accuracy. This approach has better accuracy than the one reported in  \cite{ak_arxiv} since in this paper an improved ECM model with more parameters is used.

\section{Data Preparation}
Since the proposed method relies on a supervised ML technique, there is a requirement of training data. The SoH of a cell is estimated based on the ECM parameters of Fig. \ref{fig:ecm}. Therefore, the training dataset must include capacity-based SoH and its corresponding ECM parameters. A popular format of the dataset that satisfies this requirement is if the dataset contains EIS information at multiple SoH conditions. This can be enhanced if the EIS is done at multiple SoC and temperatures for the given SoH. The proposed approach uses this dataset in the following way:
\begin{enumerate}
    \item For each SoH, choose four discrete impedances (see Section II) to create the impedance vector $\boldsymbol{Z}$ and the corresponding frequency vector $\boldsymbol{\omega}$. This process is to be repeated  at different SoC and temperature, if applicable.
    \item Compute the ECM parameters using the analytical equations shown in Fig. \ref{fig:ecm_param_extract}.
    \item Create a dataset with SoH and ECM vectors.
    \item Split the dataset into training and testing sets.
    \item Determine the coefficients of linear regression in (\ref{eq:regr_formulation}) after training.
    \item For the test data, estimate the mean absolute error (MAE) and root mean square error (RMSE) to benchmark the performance of the proposed approach.
\end{enumerate}
Note that EIS is not strictly mandatory. In this paper, EIS is described since the public datasets used for validation in this paper provide EIS vs SoH data. A training dataset could be created by directly computing discrete impedances (see Fig. \ref{fig:online_imp_calc}). This method is significantly faster than EIS and does not require specialized equipment.

\subsection{Overview of the public datasets used}
\begin{figure*}[htbp] 
    \centering 
    \begin{subfigure}[b]{0.32\textwidth} 
        \includegraphics[width=\textwidth]{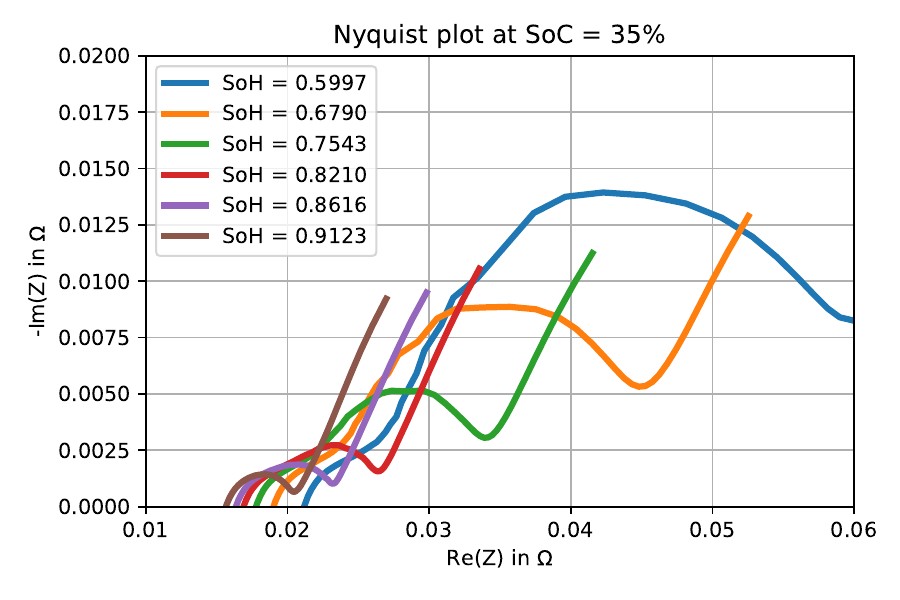}
        \caption{SoC = 35\%}
        \label{fig:nyquist_germany1} 
    \end{subfigure}
    \hfill 
    \begin{subfigure}[b]{0.32\textwidth} 
        \includegraphics[width=\textwidth]{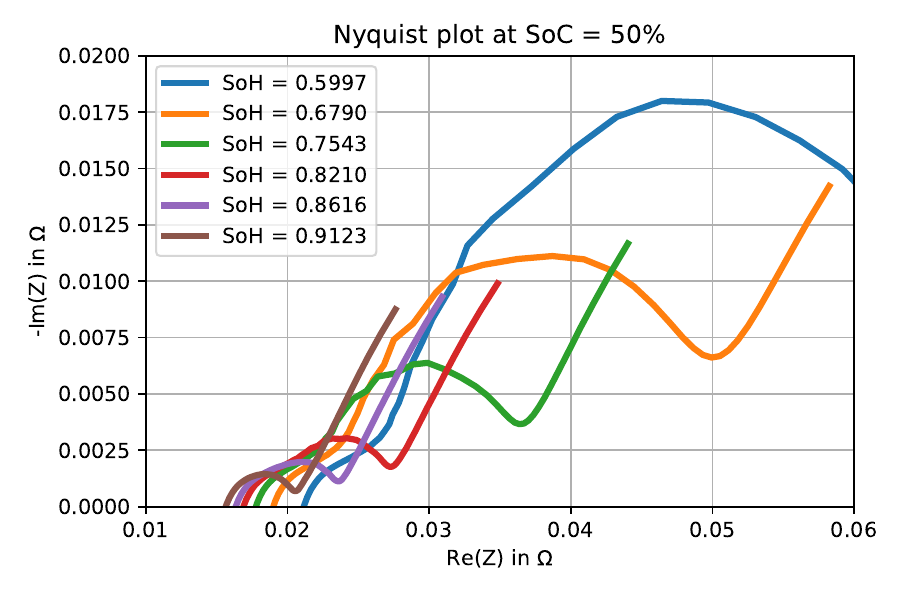}
        \caption{SoC = 50\%}
        \label{fig:nyquist_germany2}
    \end{subfigure}
    \hfill 
    \begin{subfigure}[b]{0.32\textwidth} 
        \includegraphics[width=\textwidth]{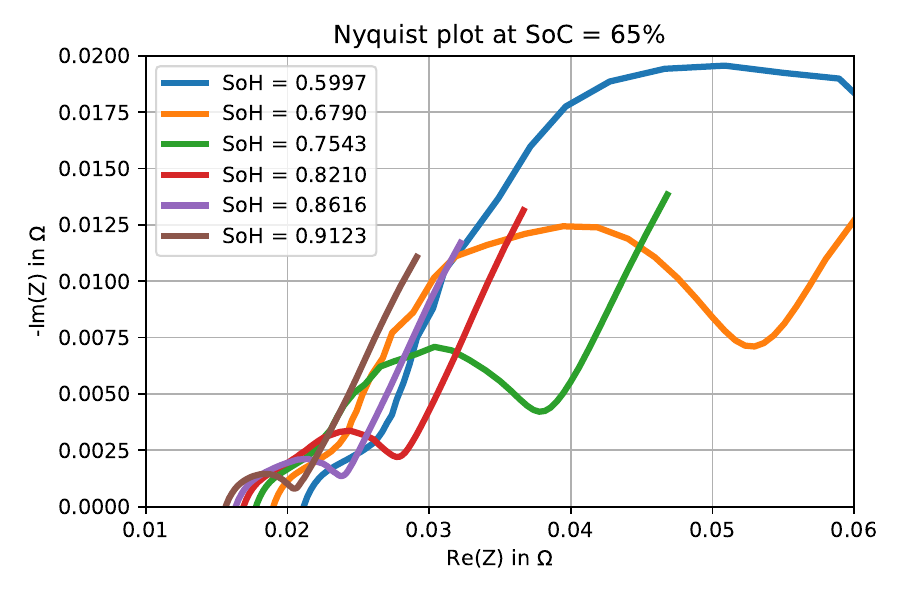}
        \caption{SoC = 65\%}
        \label{fig:nyquist_germany3}
    \end{subfigure}
    \caption{Nyquist plots at various SoH conditions from a public dataset \cite{german_dataset}.} 
    \label{fig:nyquist_germany} 
\end{figure*}

In this paper, two public datasets \cite{wmg_dataset,german_dataset} are used. Both the datasets contain EIS information at multiple values of capacity-based SoH. Table \ref{tab:dataset_summary} summarizes the key features of the two datasets.
\begin{table}[ht]
\centering
\caption{Features of the public datasets used in this paper.}
\begin{footnotesize}
\begin{tabular}{cccc}
 \toprule
Dataset & SoH data points & \makecell{Are tests done \\ at multiple SoC?} & \makecell{Are tests done \\ at multiple temperatures?} \\  \midrule  
\cite{german_dataset}  &  $>$10 &  Yes  &  No \\
 \midrule 
\cite{wmg_dataset} &  5 & Yes &  Yes \\
\bottomrule
\end{tabular}
\end{footnotesize}
\label{tab:dataset_summary}
\end{table}
Fig. \ref{fig:nyquist_germany} shows Nyquist plots at different SoH conditions and at three SoC from the dataset \cite{german_dataset}. The SoC values are 35\%, 50\% and 65\% respectively in Figs. \ref{fig:nyquist_germany1}, \ref{fig:nyquist_germany2} and \ref{fig:nyquist_germany3} respectively. 
This dataset contains a higher number of SoH breakpoints. In Fig. \ref{fig:nyquist_germany}, only six SoH breakpoints are shown for brevity. It can be observed that the behavior of the Nyquist plot is as per the expected trend where with higher aging (or lower SoH), the impedance increases. 

\begin{figure}[htbp] 
    \centering 
    \begin{subfigure}[b]{0.32\textwidth} 
        \includegraphics[width=\textwidth]{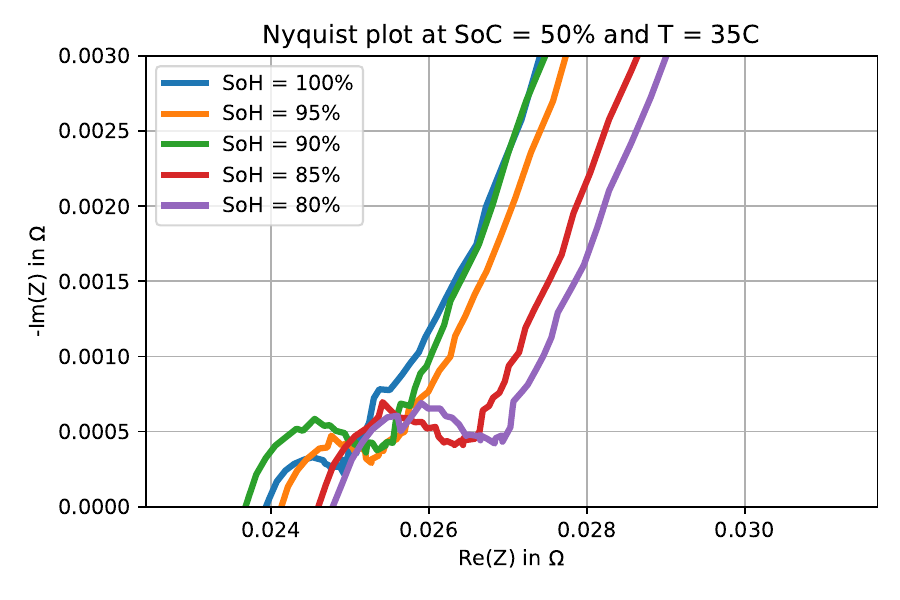}
        \caption{SoC = 50\% and temperature = $35^\circ C$}
        \label{fig:nyquist_wmg1} 
    \end{subfigure}
    \begin{subfigure}[b]{0.32\textwidth} 
        \includegraphics[width=\textwidth]{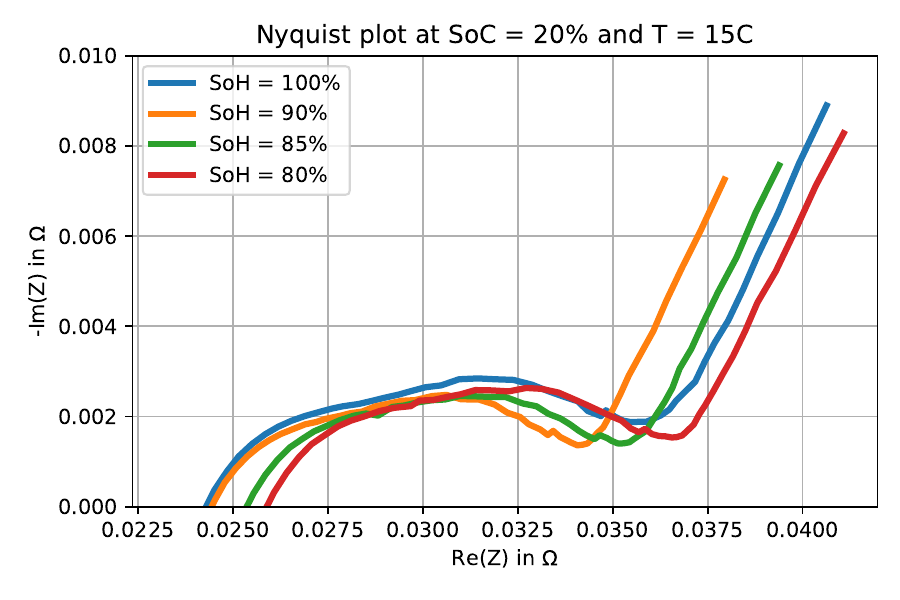}
        \caption{SoC = 20\% and temperature = $15^\circ C$}
        \label{fig:nyquist_wmg2}
    \end{subfigure}
    \caption{Nyquist plots at various SoH conditions from a public dataset \cite{wmg_dataset}.} 
    \label{fig:nyquist_wmg} 
\end{figure}

The second dataset from \cite{wmg_dataset} shows similar trend for the Nyquist plots. This is shown in Fig. \ref{fig:nyquist_wmg}. Here the dataset contains five breakpoints for the SoH as indicated in Table \ref{tab:dataset_summary}. The Nyquist plots are shown at two different conditions of cell ambient temperature and SoC. 

In Fig. \ref{fig:nyquist_wmg1}, the Nyquist plots are from an EIS test done at an SoC of 50\% and an ambient temperature of $35^\circ C$, while in Fig. \ref{fig:nyquist_wmg1}, the Nyquist plots are from an EIS test done at an SoC of 20\% and an ambient temperature of $15^\circ C$. It can also be observed that the impedance is generally higher at lower temperatures and lower SoH.

\subsection{Extracting ECM parameters from the datasets}
\begin{figure*}[htbp] 
    \centering 
    \begin{subfigure}[b]{0.32\textwidth} 
        \includegraphics[width=\textwidth]{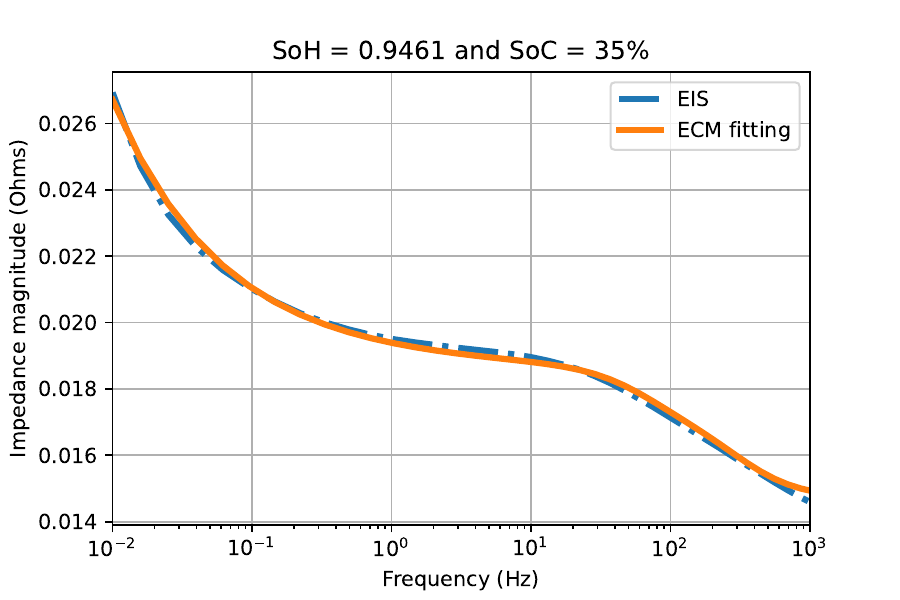}
        \caption{SoC = 35\% and SoH = 94.61\% }
        \label{fig:compare_eis_germany1} 
    \end{subfigure}
    \hfill 
    \begin{subfigure}[b]{0.32\textwidth} 
        \includegraphics[width=\textwidth]{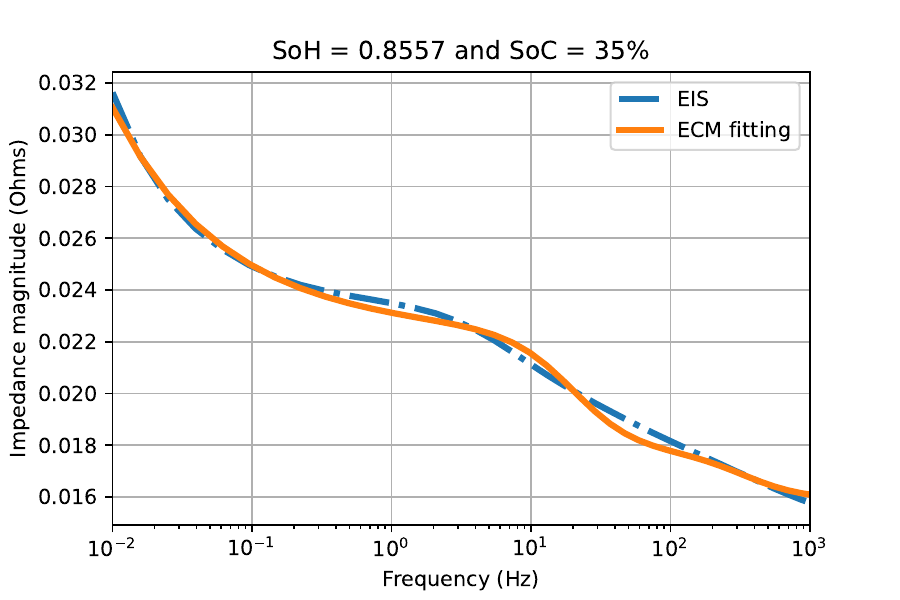}
        \caption{SoC = 35\% and SoH = 85.57\% }
        \label{fig:compare_eis_germany2}
    \end{subfigure}
    \hfill 
    \begin{subfigure}[b]{0.32\textwidth} 
        \includegraphics[width=\textwidth]{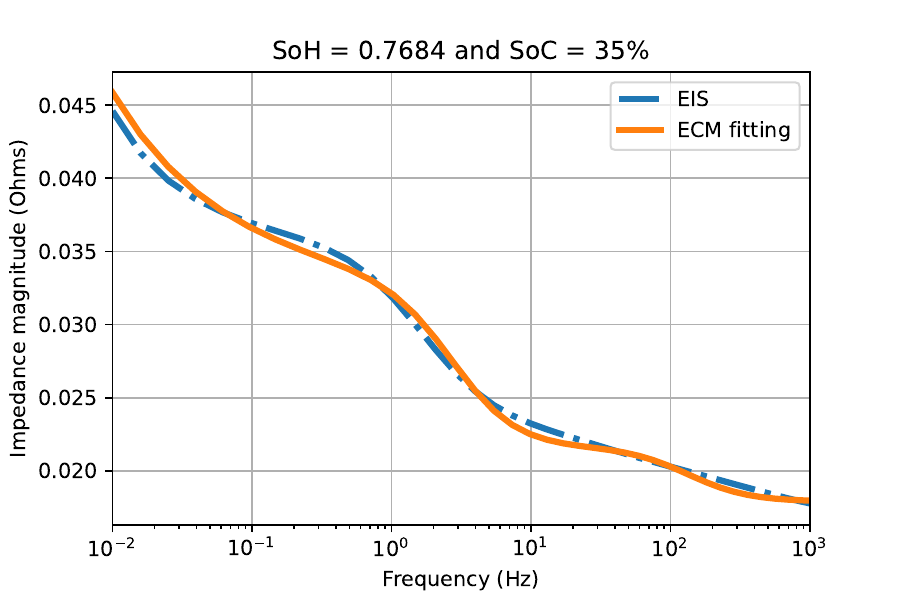}
        \caption{SoC = 35\% and SoH = 76.84\%}
        \label{fig:compare_eis_germany3}
    \end{subfigure}
    \caption{Comparing the impedance plot from proposed ECM development approach with EIS at different SoH conditions using \cite{german_dataset}.} 
    \label{fig:compare_eis_germany} 
\end{figure*}

The accuracy of the proposed method to extract the ECM parameters is illustrated by comparing the resulting impedance plot with the available impedance plot from the EIS data. Fig. \ref{fig:compare_eis_germany} shows the comparison at three different SoH conditions and at a SoC = 35\%. This is been validated for the dataset \cite{german_dataset}. Similar results are obtained for the second dataset \cite{wmg_dataset} as well.
The summary of the ECM parameters and model-fitting error is provided in Table \ref{tab:ecm_summary_error}
\begin{table}[ht]
\centering
\caption{ECM parameters corresponding to Fig. \ref{fig:compare_eis_germany}}
\begin{scriptsize}
\begin{tabular}{cccccccc}
 \toprule
\makecell{SoH \\ (\%)} & \makecell{$R_0$ \\ $(m\Omega)$} & \makecell{$R_1$ \\ $(m\Omega)$} & \makecell{$C_1$ \\ $(F)$} & \makecell{$A_w$ \\ $(m\Omega (\frac{rad}{s})^{\frac{1}{2}})$}  & \makecell{$R_2$ \\ $(m\Omega)$} & \makecell{$C_2$ \\ $(F)$} & \makecell{RMSE \\ (\%)}  \\  \midrule  
76.8& 17.9 & 11.5 & 5.2 & 4.1 & 3.7 & 0.34 & 1.9   \\  \midrule 
85.6& 15.9 & 4.7 & 1.7 & 2.7 & 1.8 & 0.27 & 1.4 \\  \midrule 
94.6& 14.7 & 1.9 & 1.2 & 2.5 & 2.1 & 0.24 & 0.9 \\  
\bottomrule
\end{tabular}
\end{scriptsize}
\label{tab:ecm_summary_error}
\end{table}

It can be observed from Table \ref{tab:ecm_summary_error} that the model error is very small ($<2\%$). The trend of parameter variation is as expected. For example, $R_0$ increases with a reduction in SoH. Thus, it can be concluded that the ECM parameter extraction provides an accurate match with the cell impedance as seen from the EIS data. Now, the ECM parameters are used as features for the ML algorithm for estimating the SoH at cell level. This is described in Section IV.  

\section{Linear Regression-Based SoH Estimation}
The two datasets used for validation in this paper differ significantly. The first dataset \cite{german_dataset} offers a substantial number of cell impedance datapoints across various states of charge (SoC) and states of health (SoH) levels, but at a fixed temperature. The second dataset \cite{wmg_dataset} provides impedance values across five different SoCs, five different SoHs, and three temperature values. Due to these differences in data availability and structure, two distinct approaches are employed for the analysis of each dataset in this study.

\subsection{Training and testing data generation}
For the dataset \cite{german_dataset}, the following approach is employed to form training and testing datasets due to the availability of a large number of data points. This dataset was developed by six institutes (labeled Inst1 through Inst6), each testing multiple commercial NMC811-silicon/graphite 18650 Li-ion cells, resulting in data from 34 cells in total. Following the recommendations in \cite{german_dataset}, data from Inst5 and Inst6 are excluded due to significant inductive distortion and increased calendar aging, respectively. 
Within each institute's data, a subset of cells is selected for training a linear regression model. This model is then tested on a different cell from the same institute that was not included in the training set. This approach allows for the assessment of the model's ability to generalize and its robustness. The specific cells chosen for training and testing are detailed in Table \ref{tab:germany_train_test}. All the EIS tests in this dataset are performed at a temperature of $23^\circ C$.
\begin{table}[ht]
\centering
\caption{Training and testing data generation from \cite{german_dataset}.}
\begin{footnotesize}
\begin{tabular}{ccc}
 \toprule
Institute &  \makecell{Cell \# for training} & \makecell{Cell \# for testing} \\  \midrule  
Inst1  &  04, 13, 14, 19 & 09 \\ \midrule 
Inst2(a)  &  06, 07 & 10 \\ \midrule 
Inst2(b)  &  11, 12  & 13 \\ \midrule 
Inst3  &  01, 02, 07, 08 & 03 \\ 
\bottomrule
\end{tabular}
\end{footnotesize}
\label{tab:germany_train_test}
\end{table}

For the dataset \cite{wmg_dataset}, three combined datasets are created, each containing EIS data at five SoH and SoC levels for a specific temperature: $15^\circ C$, $25^\circ C$ and $35^\circ C$. This dataset uses cylindrical Li-ion cells of 21700 geometry with NMC811 chemistry.
Training is conducted separately for each temperature condition. Within each temperature dataset, 60\% of the data is allocated for training and 40\% for testing. The assumption of constant temperature during training and testing is based on the BMS's ability to perform SoH estimation at a given temperature, thus avoiding the need to account for temperature variations within the model.
While incorporating temperature variations into the training would be feasible with a larger dataset, the limited availability of data at only three temperature points justifies this approach in this paper.

\subsection{Results}
\begin{figure*}[htbp]  
    \centering      

    \begin{subfigure}{0.4\textwidth}  
        \includegraphics[width=\textwidth]{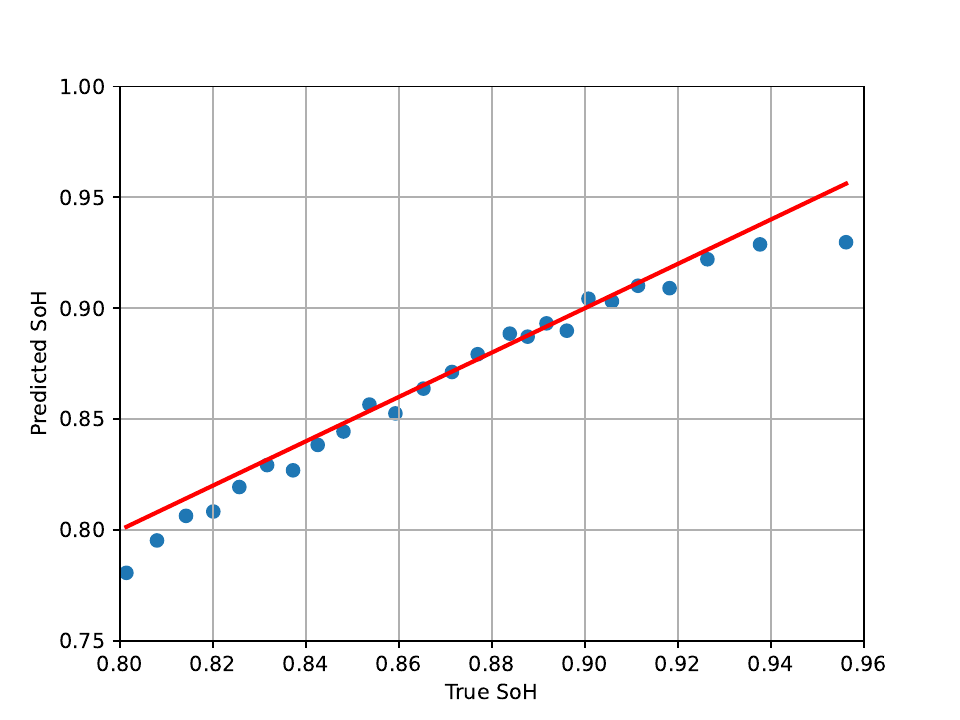}
        \caption{Results for cells from Inst1}
        \label{germany_results_low_soc1}
    \end{subfigure}
    \begin{subfigure}{0.4\textwidth}  
        \includegraphics[width=\textwidth]{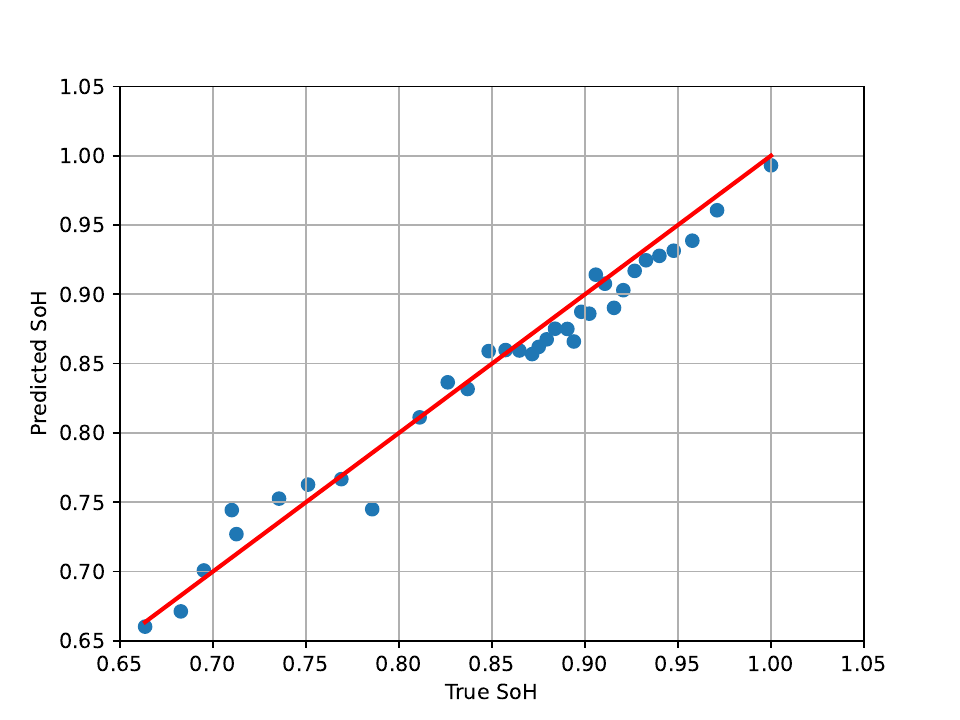}
        \caption{Results for cells from Inst2, group (a)}
        \label{germany_results_low_soc2}
    \end{subfigure}

    \medskip  

    \begin{subfigure}{0.4\textwidth}  
        \includegraphics[width=\textwidth]{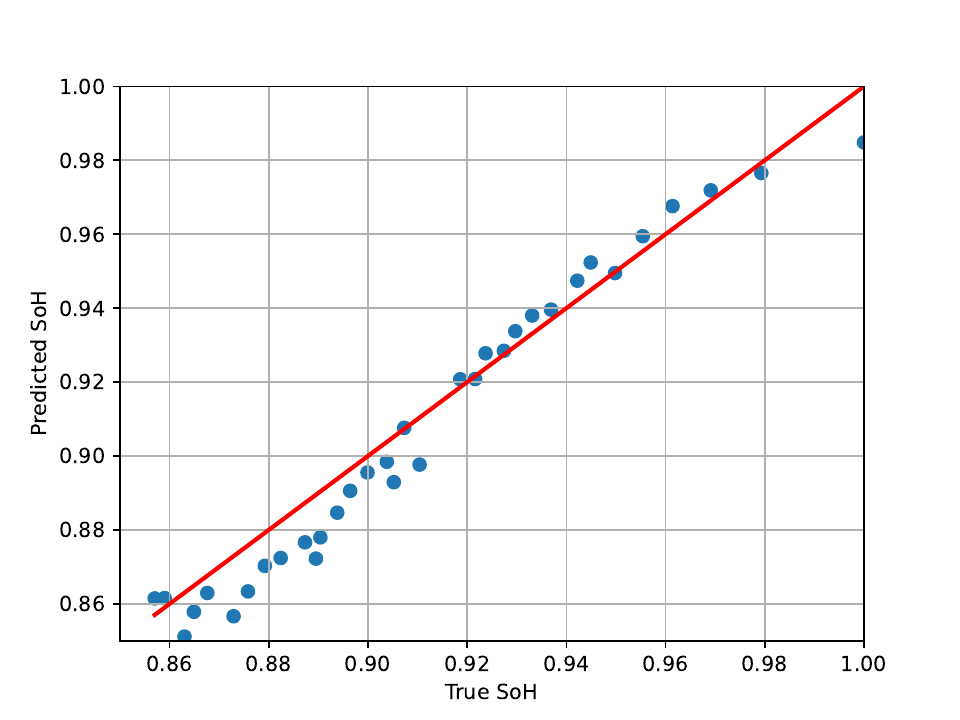}
        \caption{Results for cells from Inst2, group (b)}
        \label{germany_results_low_soc3}
    \end{subfigure}
    \begin{subfigure}{0.4\textwidth}  
        \includegraphics[width=\textwidth]{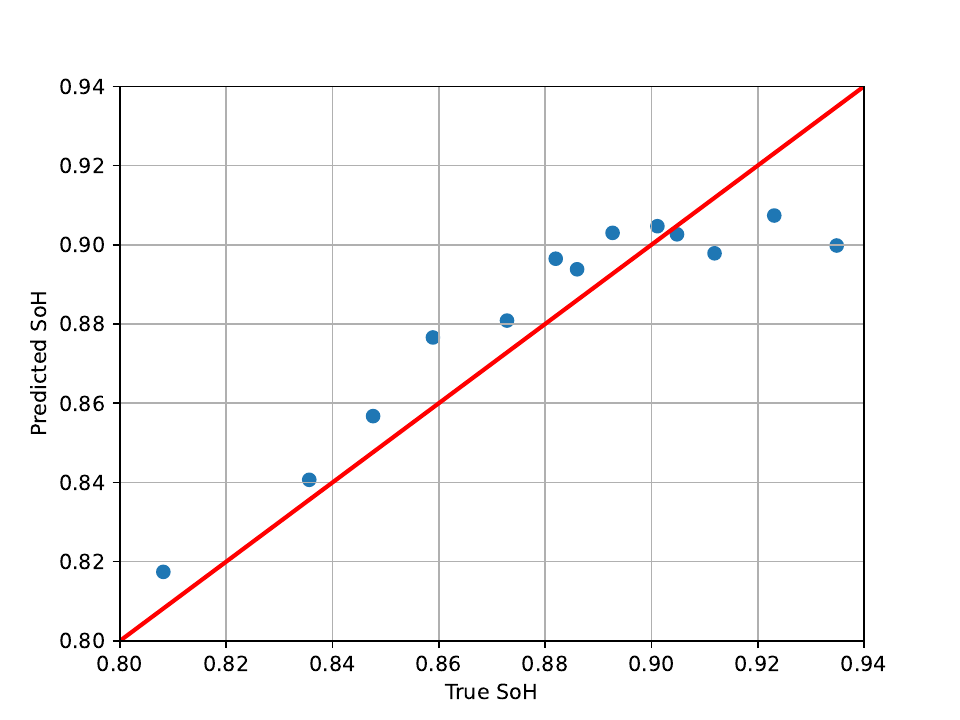}
        \caption{Results for cells from Inst3}
        \label{germany_results_low_soc4}
    \end{subfigure}

    \caption{Comparison of estimated SoH with true SoH for different datasets from \cite{german_dataset}}
    \label{fig:germany_results_low_soc}  
\end{figure*}

Once the training and testing datasets are created as described in Section IV-A, linear regression is applied to determine the coefficients in (\ref{eq:regr_formulation}).
Fig. \ref{fig:germany_results_low_soc} show the results of the proposed SoH estimation method using linear regression. The cell data as per Table \ref{tab:germany_train_test} is trained at low SoC (20\%). The results for the four datasets indicate that there is a good linear fit and linear regression approach followed in this work gives accurate results. The mean absolute error (MAE) for the four cases is determined to be less than 2\%. Similar results are obtained at other SoCs. It will be the choice of the BMS to perform the SoH estimation at any predetermined SoC level.

The summary of model fitness score $R^2$, MAE and RMSE is provided in Table \ref{tab:summary_mae_german}. It can be observed that RMSE is higher than MAE since outliers will have a bigger influence in the RMSE compared to MAE. Overall, it can be concluded that the results show an excellent fit with very low estimation errors.
\begin{table}[ht]
\centering
\caption{SoH estimation performance summary for \cite{german_dataset}.}
\begin{footnotesize}
\begin{tabular}{cccc}
 \toprule
Data source &  \makecell{$R^2$} & \makecell{MAE (\%)} & \makecell{RMSE (\%)}\\  \midrule  
Inst1  &  0.954 & 0.65 & 0.89 \\ \midrule 
Inst2(a)  &  0.970 & 1.27 & 1.54 \\ \midrule 
Inst2(b)  &  0.948 & 0.68 & 0.83 \\ \midrule 
Inst3  &  0.968 & 2.00 & 3.05 \\ 
\bottomrule
\end{tabular}
\end{footnotesize}
\label{tab:summary_mae_german}
\end{table}

\begin{figure*}[htbp] 
    \centering 
    \begin{subfigure}[b]{0.325\textwidth} 
        \includegraphics[width=\textwidth]{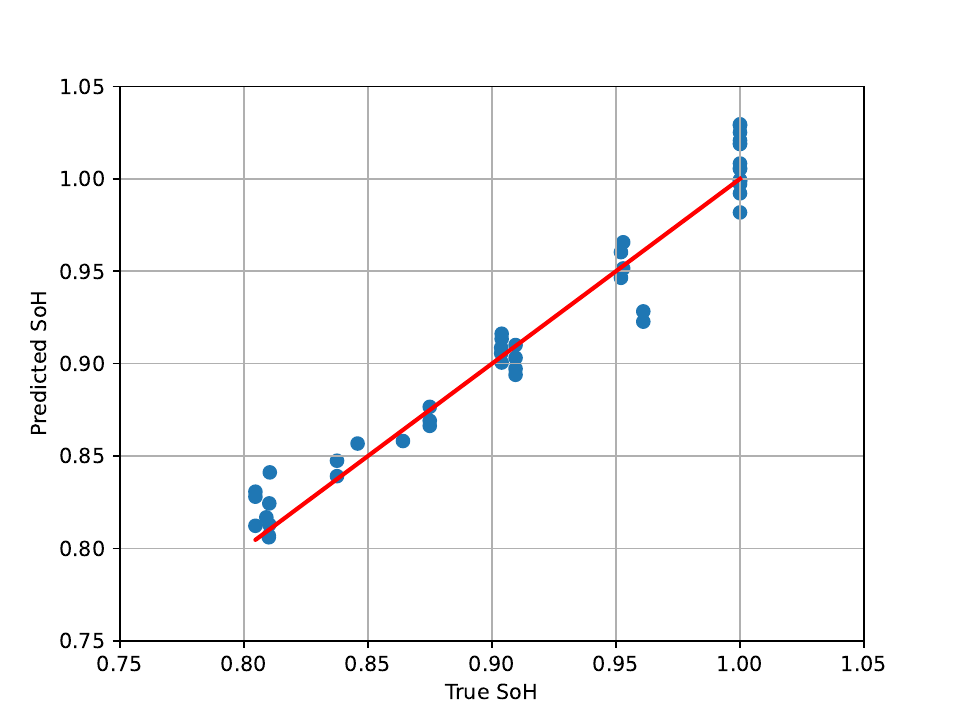}
        \caption{Results for cells trained at $15^\circ C$.}
        \label{wmgt15} 
    \end{subfigure}
    \hfill 
    \begin{subfigure}[b]{0.325\textwidth} 
        \includegraphics[width=\textwidth]{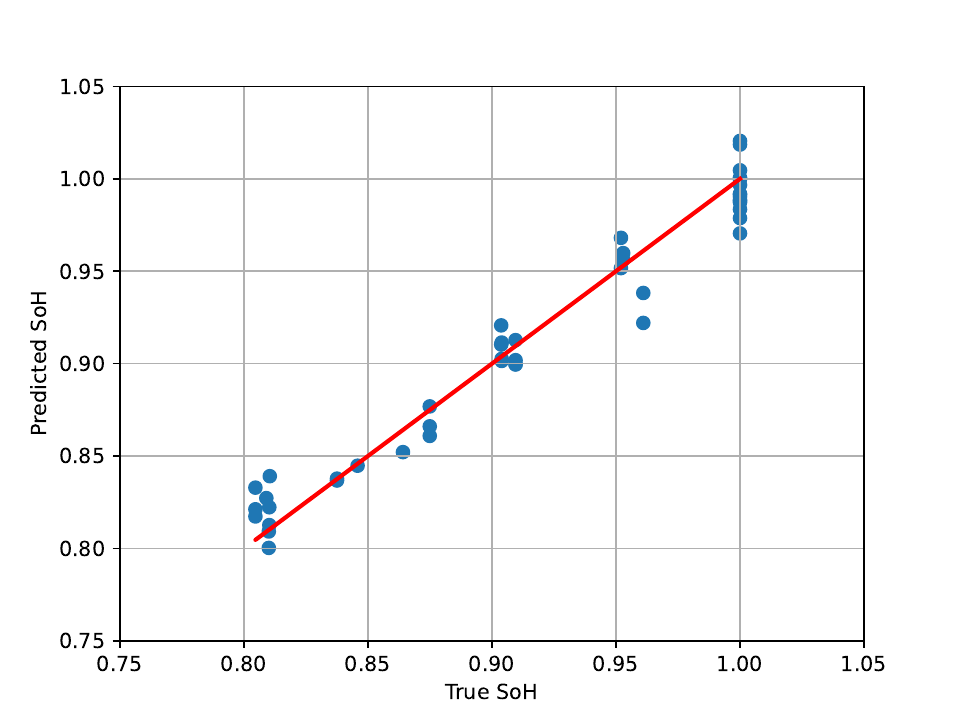}
        \caption{Results for cells trained at $25^\circ C$.}
        \label{wmgt25}
    \end{subfigure}
    \begin{subfigure}[b]{0.325\textwidth} 
        \includegraphics[width=\textwidth]{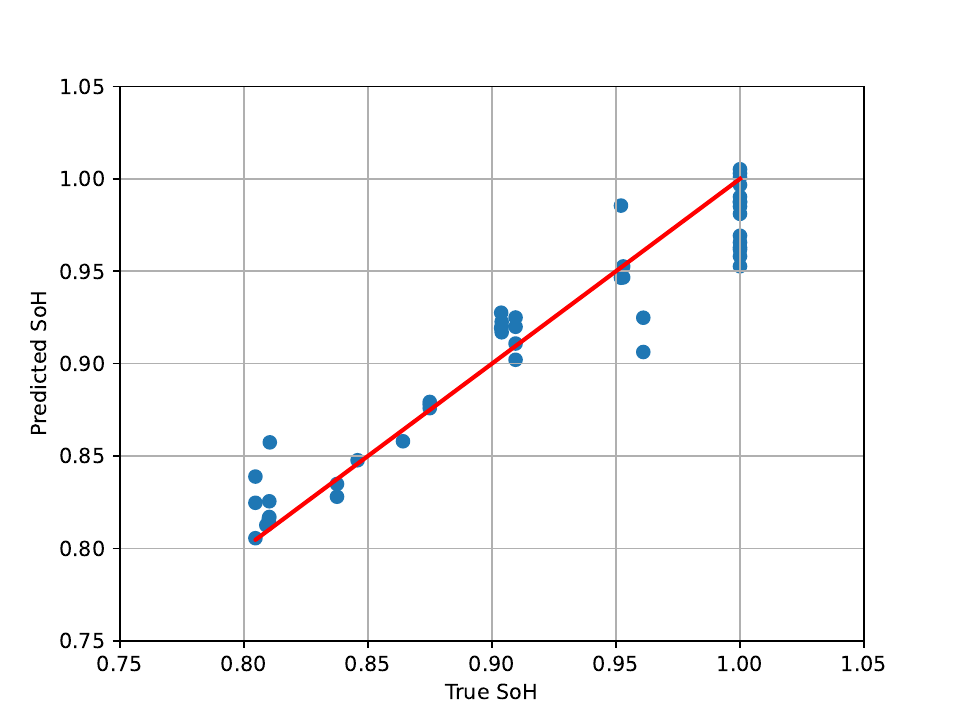}
        \caption{Results for cells trained at $35^\circ C$.}
        \label{wmgt35} 
    \end{subfigure}   
    \caption{Comparison of estimated SoH with true SoH for different datasets from \cite{wmg_dataset}} 
    \label{fig:wmg_results} 
\end{figure*}

Similar performance is observed for the proposed approach using the second dataset \cite{wmg_dataset}. Fig. \ref{fig:wmg_results}.
The performance indicators $R^2$, MAE and RMSE are determined for this dataset and summarized in Table \ref{tab:summary_mae_wmg}.
\begin{table}[ht]
\centering
\caption{SoH estimation performance summary for \cite{wmg_dataset}.}
\begin{footnotesize}
\begin{tabular}{cccc}
 \toprule
Data source &  \makecell{$R^2$} & \makecell{MAE (\%)} & \makecell{RMSE (\%)}\\  \midrule  
Data at  $15^\circ C$  &  0.958 & 1.14 & 1.51 \\ \midrule 
Data at  $25^\circ C$ &  0.963 & 1.01 & 1.41 \\ \midrule 
Data at  $35^\circ C$  &  0.911 & 1.60 & 2.18 \\ 
\bottomrule
\end{tabular}
\end{footnotesize}
\label{tab:summary_mae_wmg}
\end{table}

As it can be observed from Tables \ref{tab:summary_mae_german} and \ref{tab:summary_mae_wmg}, the proposed method results in a maximum MAE of just 2\%. The main advantages of the proposed approach are summarized as follows:
\begin{itemize}
    \item The ML algorithm's features consist of ECM parameters, calculated using four discrete impedance measurements and a system of algebraic equations.
    \item The feature extraction process avoids computationally intensive methods such as Fourier transforms or nonlinear optimization.
    \item SoH estimation is performed using a linear regression model fit to the extracted features. This approach offers a lightweight alternative to neural-network-based methods, which often require significant memory and nonlinear activation functions.
\end{itemize}

\section{Conclusion}
AI/ML methods are increasingly employed for accurate SoH estimation in Li-ion batteries. However, the complexity and resource demands of many of these methods negatively impact their real-world implementation in commercial BMS processors. This paper addresses this challenge by introducing a low-complexity, data-efficient solution for Li-ion battery SoH estimation. The proposed method utilizes four discrete impedance values, easily obtainable online, as inputs. These are transformed into ECM parameters via algebraic equations, which subsequently feed into a linear regression model. Validation across two distinct datasets demonstrates exceptional model fit, with high R-squared values and low MAE ($<2\%$). This performance highlights the proposed method's suitability for implementation in commercial BMS processors, offering an additional SoH estimation technique, enhancing accuracy and redundancy - critical factors for vehicle safety in e-mobility applications.

\appendices

\section{Deriving analytical expressions for ECM parameters}
If impedance values are available at four well separated frequencies as shown in Fig. \ref{fig:freq_sel}, it is possible to derive analytical equations that relate the parameters of the ECM in Fig. \ref{fig:ecm} to the real and complex components of the four impedances and their corresponding frequencies. The final expressions are already shown in Fig. \ref{fig:ecm_param_extract}.

\begin{figure}[htbp]  
\centering  
\begin{subfigure}[c]{0.35\linewidth} 
    \includegraphics[width=\textwidth]{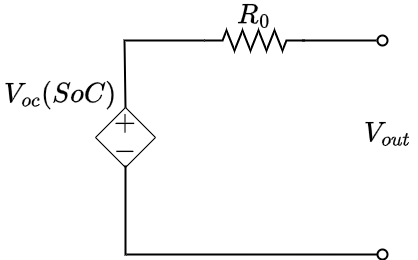}  
    \caption{\centering{$\omega=\omega_{high}$ and $Z_{eq}=Z_{fhigh}$}}  
    \label{fig:reduced_ecms1}
\end{subfigure}
\hfill 
\begin{subfigure}[c]{0.55\linewidth} 
    \includegraphics[width=\textwidth]{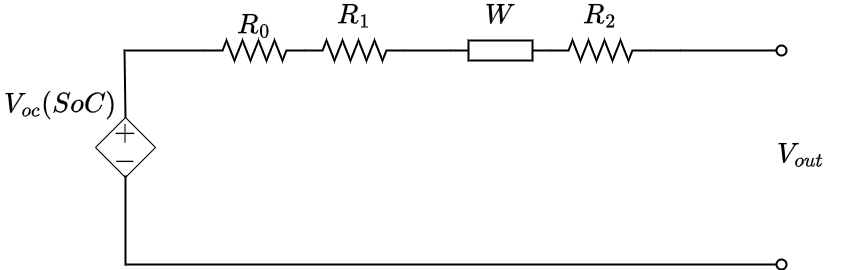} 
    \caption{\centering{$\omega=\omega_{low}$ and $Z_{eq}=Z_{flow}$}}
    \label{fig:reduced_ecms2}
\end{subfigure}

\medskip 

\begin{subfigure}[c]{0.49\linewidth} 
    \includegraphics[width=\textwidth]{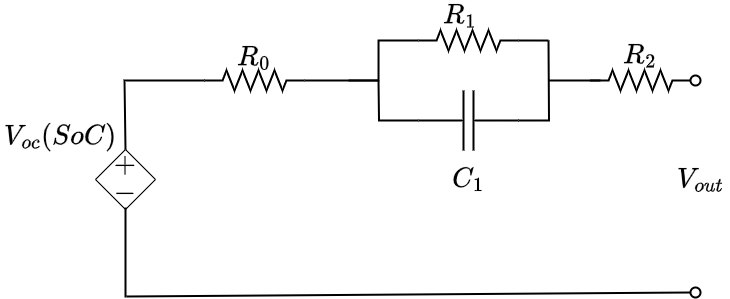} 
    \caption{\centering{$\omega=\omega_{mid1}$ and $Z_{eq}=Z_{fmid1}$}}
    \label{fig:reduced_ecms3}
\end{subfigure}
\hfill
\begin{subfigure}[c]{0.49\linewidth} 
    \includegraphics[width=\textwidth]{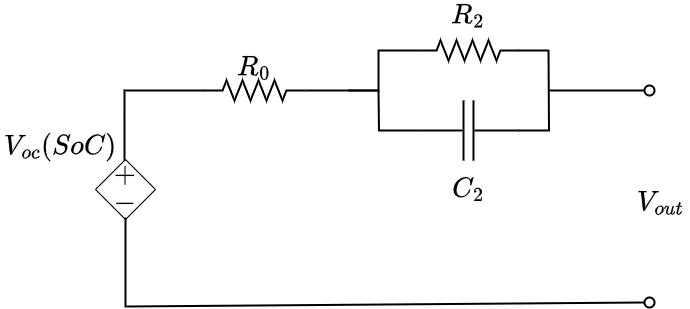}
    \caption{\centering{$\omega=\omega_{mid2}$ and $Z_{eq}=Z_{fmid2}$}}
    \label{fig:reduced_ecms4}
\end{subfigure}
\caption{Simplified battery equivalent circuits at different frequencies of interest.} 
\label{fig:reduced_ecms}
\end{figure}

For each frequency in the frequency vector in (\ref{eq:freq_vector}), the ECM can be approximated to a simpler circuit considering the impact of frequency on capacitive and Warburg impedance. These equivalent circuits are shown in Fig. \ref{fig:reduced_ecms} corresponding to each frequency element under consideration.
For example, when the frequency is high, the capacitive impedances become very small and hence the equivalent impedance reduces to just $R_0$ as seen in Fig. \ref{fig:reduced_ecms1}. When the frequency is very low, the capacitive impedances can be considered as open circuit, resulting in Fig. \ref{fig:reduced_ecms2}. Similarly, for the the middle frequency range when $\omega=\omega_{mid1}$, the impedance from capacitance $C_2$ is still considered to act as open circuit resulting in a simplified ECM shown in Fig. \ref{fig:reduced_ecms3}. Here the frequency is high enough that the Warburg impedance becomes negligible but the impedance due to $C_1$ is significant. Finally, when $\omega=\omega_{mid2}$ with $\omega_{mid2}>\omega_{mid1}$, the ECM reduces just to a parallel combination of $R_2$, $C_2$ in series with $R_0$. This is shown in Fig. \ref{fig:reduced_ecms4}.

Note that the impedance vector in (\ref{eq:imp_vector}) is expressed in terms of the real and imaginary parts of the impedance as follows.
\begin{align}
 \boldsymbol{Z} =& [R_{fhigh},~X_{fhigh},~R_{fmid1},~X_{fmid1}, \notag \\ 
 &R_{fmid2},~X_{fmid2},~R_{flow},~X_{flow},]^T \label{eq:imp_vector_expand}   
\end{align}
Now, with the help of the simplified ECMs and algebraic manipulations similar to the ones done in \cite{ak_arxiv} for a simplified model, the parameters of the ECM are determined as follows.   
\begin{align}
  R_0 &= R_{fhigh} \label{eq:R0} \\
  A_w &= X_{flow}\sqrt{2\omega_{low}} \label{eq:Aw} \\
  C_1 &= \frac{X_{fmid1}}{\omega_{mid1}(R_{fmid1}-R_0)(R_{flow}-R_0-X_{flow})} \label{eq:C1} \\
  R_2 &= (R_{fmid2}-R_0)\left[1+ \left(\frac{X_{fmid2}}{R_{fmid2}-R_0}\right)^2 \right] \label{eq:R2} \\
  C_2 &= \frac{X_{fmid2}}{\omega_{mid2}(R_{fmid2}-R_0)^2 \left[1+ \left(\frac{X_{fmid2}}{R_{fmid2}-R_0}\right)^2 \right]} \label{eq:C2} \\
  R_1 &= R_{flow}-R_0-X_{flow} - R_2 \label{eq:R1}
\end{align}

It can be observed from (\ref{eq:R0}) -- (\ref{eq:R1}), that the knowledge of the impedance vector (\ref{eq:imp_vector_expand}) and the frequency vector (\ref{eq:freq_vector}) is sufficient to extract all the ECM parameters. The accuracy of the model thus developed has been validated as shown in Fig. \ref{fig:compare_eis_germany} and Table \ref{tab:ecm_summary_error}.


\end{document}